# Comment to "Comment on 'Unified treatment for the evaluation of arbitrary multielectron multicenter molecular integrals over Slater-type orbitals with noninteger principal quantum numbers'"

Telhat Özdogan[*]

*Department of Physics, Rize Faculty of Arts and Sciences, Karadeniz Technical University, Rize, TURKEY*

**Abstract**

In a recent publication (Telhat Özdogan, Int. J. Quantum Chem., 92 (2003) 419), we presented a unified algorithm for the evaluation of multicenter multielectron integrals over Slater type orbitals with noninteger principal quantum numbers, using the one center expansion formula for Slater type orbitals with integer principal quantum numbers (E. Öztekin et. all., J. Mol. Struct. Theochem, **544** (2001) 69; I.I. Guseinov et all., Z. Struct. Khim., 23 (1987) 148 (in Russian)). Guseinov in his comment (arXiv-physics/0510235) on our publication claims that the presented formulae in our paper could be derived from formulae in his papers by simple algebra. We give a brief response to some of his claims wherever scientifically appropriate. We also assert that the comments are not scientifically justified and rhetoric is highly inappropriate. It should be noted also that it is not easy to understand that what Guseinov intends to do by transforming our procedure into his procedure, which we don't evaluate as a development but can be regarded as a nice sound of science.

**KEY WORDS:** Slater-type orbital, nuclear attraction integral, Legendre function.

**AMS subject classification:** 81-08,81-V55,81V45

In the comment of I.I. Guseinov [1], it is claimed that the formulae we presented in our recent paper [2] are derived from the relationships in his studies [3], by changing the summation indices.

As can be seen from our paper [2] that the early work of Guseinov [4] (cited as Ref. 19 in our paper [2]) is cited in constructing our procedure. It is understood that Guseinov do not read our papers, He just write comments as he always does.

The constructed formulae in our recent paper [2] is used in the evaluation of arbitrary multielectron multicenter molecular integrals over Slater type orbitals with noninteger principal quantum numbers and can be understand clearly. That is, the expressions for multielectron integrals over Slater type orbitals with noninteger principal quantum numbers have been obtained independently, not by changing the summation indices of the formulae in papers by Guseinov. The efficiency and accuracy of our algorithm is tested for a wide range of molecular parameters. It is seen that the algorithm presented here for the calculation of arbitrary multielectron multicenter molecular integrals for noninteger *n*-Slater type orbitals is the first in the literature, efficient and can be used in molecular structure calculations without loss of significant figures as occur in some weak methods, especially by Guseinov [5]. It is advised to read the excellent paper of Barnett [6] for digital erosion may occur in methods by Guseinov [5] and others.

---
[*] E-mail: telhat@gmail.com

Guseinov causes misunderstanding for readers and publishers by citing his comments but hiding our replies. He should be warned by readers and publishers for this ethical problem. Therefore, we advise to see our replies to his unjust comments in the link below: http://www.geocities.com/telhat/telhatozdogan.html. In this manner, we advise readers to communicate with me!!!